\documentclass[multphys,vecphys]{svmult}

\usepackage{makeidx}         % allows index generation
\usepackage{graphicx}        % standard LaTeX graphics tool
\usepackage{multicol}        % used for the two-column index
\usepackage[bottom]{footmisc}% places footnotes at page bottom
\makeindex             % used for the subject index
\begin{document}

\title*{{\it XMM-Newton} view of MS0735+7421: the most energetic AGN outburst in a galaxy cluster}
\author{M. Gitti \inst{1}
\and B. R. McNamara \inst{1,2}
\and P. E. J. Nulsen \inst{3}
\and M. W. Wise \inst{4}}
\institute{Dept. of Physics and Astronomy, Ohio University,
Clippinger Labs, Athens, OH 45701 (USA)
\texttt{gitti@phy.ohiou.edu}
\and University of Waterloo,
200 University Avenue West,
Waterloo, Ontario N2L 3G1 (Canada)
\texttt{mcnamara@uwaterloo.ca}
\and Harvard-Smithsonian Center for Astrophysics,
60 Garden Street, Cambridge, MA 02138 (USA)
\texttt{pnulsen@head.cfa.harvard.edu}
\and Astronomical Institute, University of Amsterdam, Kruislaan 403,
1098 SJ Amsterdam (The Netherlands)
\texttt{wise@science.uva.nl}}
\titlerunning{MS0735+7421: the most energetic AGN outburst in a galaxy cluster}
\authorrunning{Gitti et al.} 
\maketitle

%%%%%%%%%%%%%%%%%%%%%%%%%%%%%%%%%%%%%%%%%%%%%%%%%%%%%%%%%%%%%%%%%%%%%%%%%%%%%%

We discuss the possible cosmological effects of powerful AGN outbursts in
galaxy clusters by starting from the results of an \textit{XMM-Newton} 
observation of the supercavity cluster MS0735+7421.

%%%%%%%%%%%%%%%%%%%%%%%%%%%%%%%%%%%%%%%%%%%%%%%%%%%%%%%%%%%%%%%%%%%%%%%%%%%%%%

\section{Introduction}
\label{intro.sec}
% Always give a unique label
% and use \ref{<label>} for cross-references
% and \cite{<label>} for bibliographic references
% use \sectionmark{}
% to alter or adjust the section heading in the running head

The majority of cooling flow clusters contain powerful radio sources 
associated with the central cD galaxies \cite{burns90}. 
As indicated by high resolution X-ray images, these radio sources have a 
profound impact on the intra-cluster medium (ICM) -- the radio lobes 
displace the X-ray emitting gas, creating X-ray deficient cavities 
(\cite{mcnamara00}, \cite{blanton03}, \cite{gitti06}).
The recent discovery of giant cavities and associated large-scale shocks 
in three galaxy clusters (MS0735+7241 \index{MS0735+7421} \cite{mcnamara05}, 
Hercules A \index{Hercules A} \cite{nulsen05a}, Hydra A \index{Hydra A}
\cite{nulsen05b}) has shown that AGN outbursts can not only affect the 
central regions, but also have an impact on cluster-wide scales.

This new development may have significant consequences for several fundamental
 problems in astrophysics.
The non-gravitational heating supplied by AGNs could represent an 
important contribution to the extra energy necessary to "pre-heat" galaxy
 clusters, thus explaining the steepening of the observed luminosity vs. 
temperature relation with respect to theoretical predictions that include
 gravity alone (\cite{markevitch98}, \cite{wu00}).
Powerful central AGN outbursts may also affect the general properties of 
the ICM (e.g., temperature and metallicity profiles, X-ray luminosity, 
gas mass fraction).
It is essential to understand well the ICM physics and evaluate the
potential impact of AGN outbursts on the mass vs. temperature ($M$-$T$) and
luminosity vs. temperature ($L$-$T$) relations, which are the foundation to 
construct the cluster mass function and use galaxy clusters as 
cosmological probes.

We address these problems by studying the X-ray properties of the most 
energetic outburst known in a galaxy cluster.
MS0735$+$7421 (hereafter MS0735) is at a redshift of 
0.216.
With $H_0 = 70 \mbox{ km s}^{-1} \mbox{ Mpc}^{-1}$, 
$\Omega_M = 1-\Omega_{\Lambda} = 0.3$, the angular scale is 3.5 kpc per 
arcsec.

%%%%%%%%%%%%%%%%%%%%%%%%%%%%%%%%%%%%%%%%%%%%%%%%%%%%%%%%%%%%%%%%%%%%%%%%%%%%

\section{{\it XMM-Newton} view of MS0735+7421}
\label{results.sec}

MS0735 was observed by \textit{XMM--Newton} in April 2005 for a total 
clean exposure time of about 50 ksec.
The X-ray image of the central region of the cluster 
(see Fig. \ref{fig:1}, top) shows twin giant cavities having 
$\sim$ 200 kpc 
diameter each, in agreement with results from the previous {\it Chandra} 
observation \cite{mcnamara05}. 
 
The presence of the radio source at the position coincident with the 
holes in the X-ray emission \cite{mcnamara05} implies that the cavities 
are filled with a population of relativistic electrons radiating at low 
radio frequencies.
The cavities may also be filled with a shock-heated thermal gas 
that can contribute to the internal pressure necessary to sustain them. 
In order to investigate this hypothesis we performed a detailed spectral 
analysis by modeling the spectra extracted in the cavity regions 
(see Fig. \ref{fig:1}, top) as the sum of ambient cluster emission and a
hot thermal plasma, each with a characteristic temperature.
Hints of a second thermal component with $kT > 10$ keV were found in the 
northern cavity, in agreement with similar estimates in other clusters
(\cite{blanton03} , \cite{fabian02}, \cite{mazzotta04}).

We also attempted a study of the shock properties.
{\it Chandra} observations reveal a feature in the X-ray surface 
brightness that has been interpreted as a weak cocoon shock driven by 
the expansion of the radio lobes that inflate the cavities \cite{mcnamara05}.
Our results from a spectral analysis of the pre-shocked and post-shocked 
gas are consistent with {\it Chandra} ones, although {\it XMM--Newton}'s
spatial resolution is too poor to unambiguously measure the temperature 
across the shock.

%%%%%%%%%%%%%%%%%%%%%%%%%%%%%%%%%%%%%%%%%%%%%%%%%%%%%%%%%%%%%%%%%%%%%%%%%%%%%%

\section{Do cavities affect the average cluster properties?}
\label{discussion.sec}

From the {\it XMM--Newton} observation of MS0735 we can get new insights 
to evaluate the impact of energetic AGN explosions on the general 
cluster properties and scaling relations, which are fundamental to use 
galaxy clusters as cosmological probes.
The main results can be summarized as follows (see \cite{gittinew} for a 
more detailed discussion):

\begin{itemize}

\item
The total energy in cavities and shock is $\sim 6 \times 10^{61}$ erg, 
making MS0735 the most energetic AGN outburst known so far.
This energy is enough to quench the nominal 
$\sim 260 \, {\rm M}_{\odot} {\rm yr}^{-1}$ cooling flow and, since most of 
the energy is deposited outside the cooling region ($\sim$ 100 kpc),
to heat the gas within 1 Mpc by $\sim$ 1/4 keV per particle. 
It thus contributes a substantial fraction of the 1 to 3 keV per particle of 
excess energy required to preheat the cluster \cite{wu00}. 
\\

\begin{figure}[ht]
\includegraphics{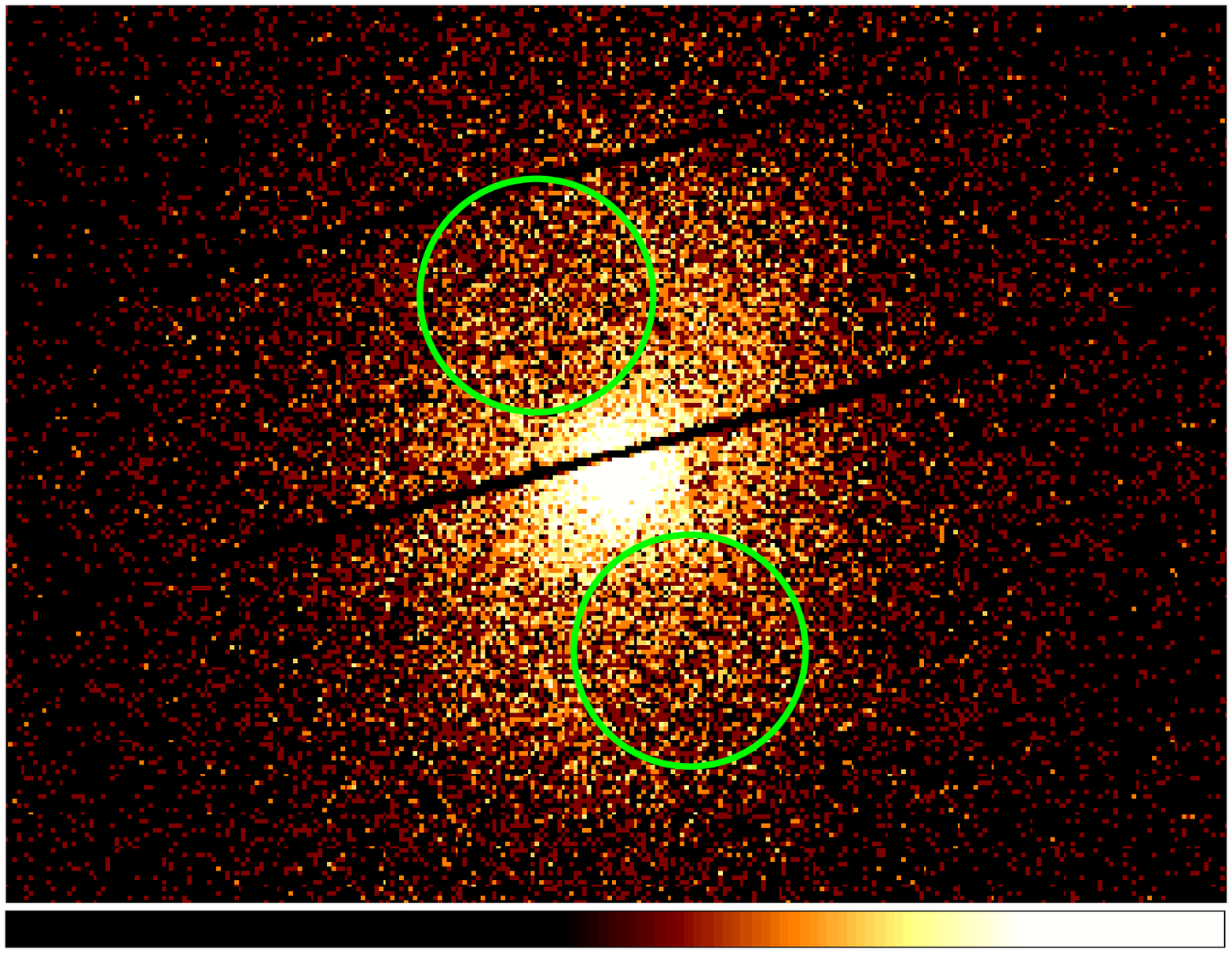}
\includegraphics{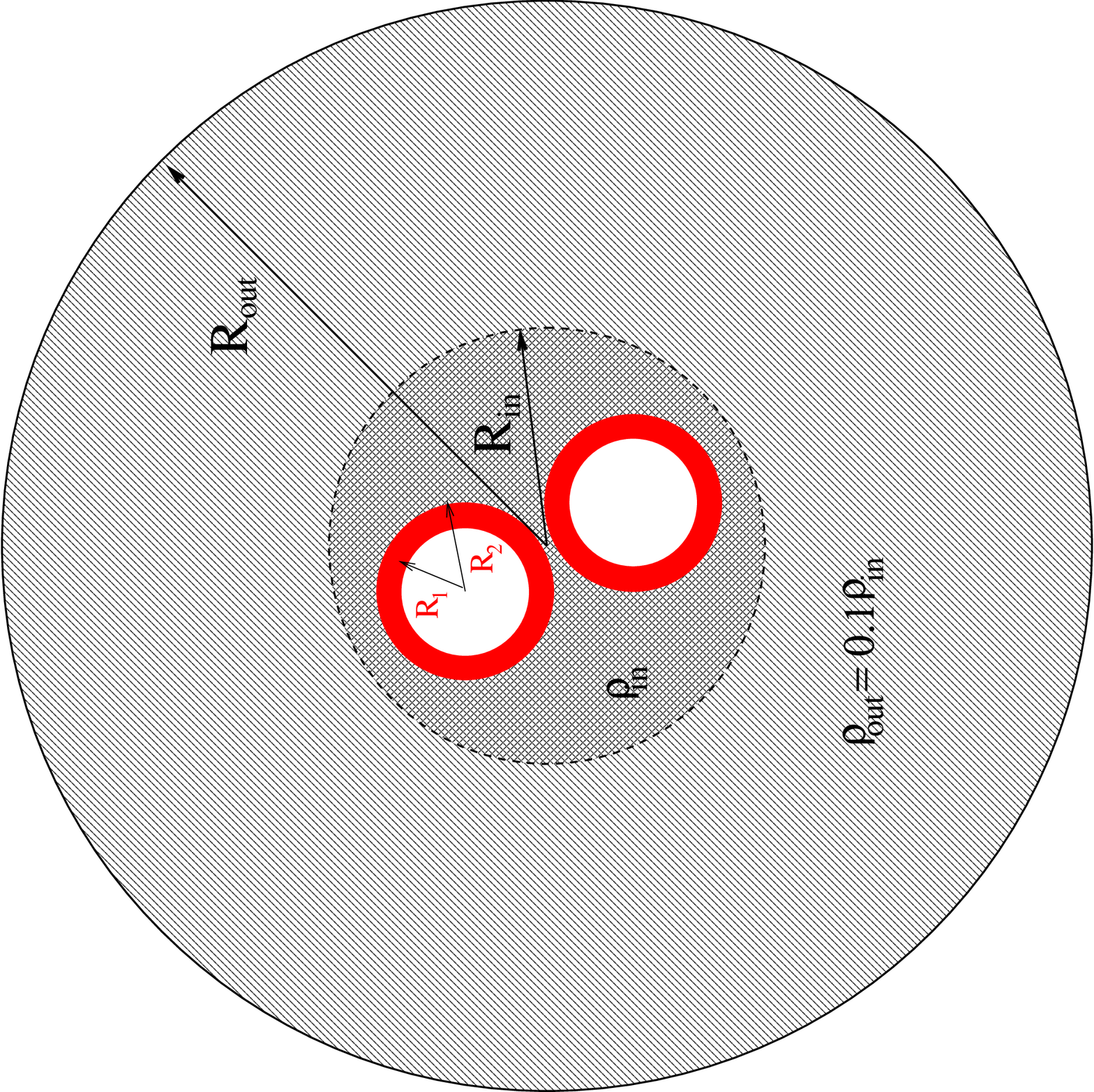}
\includegraphics{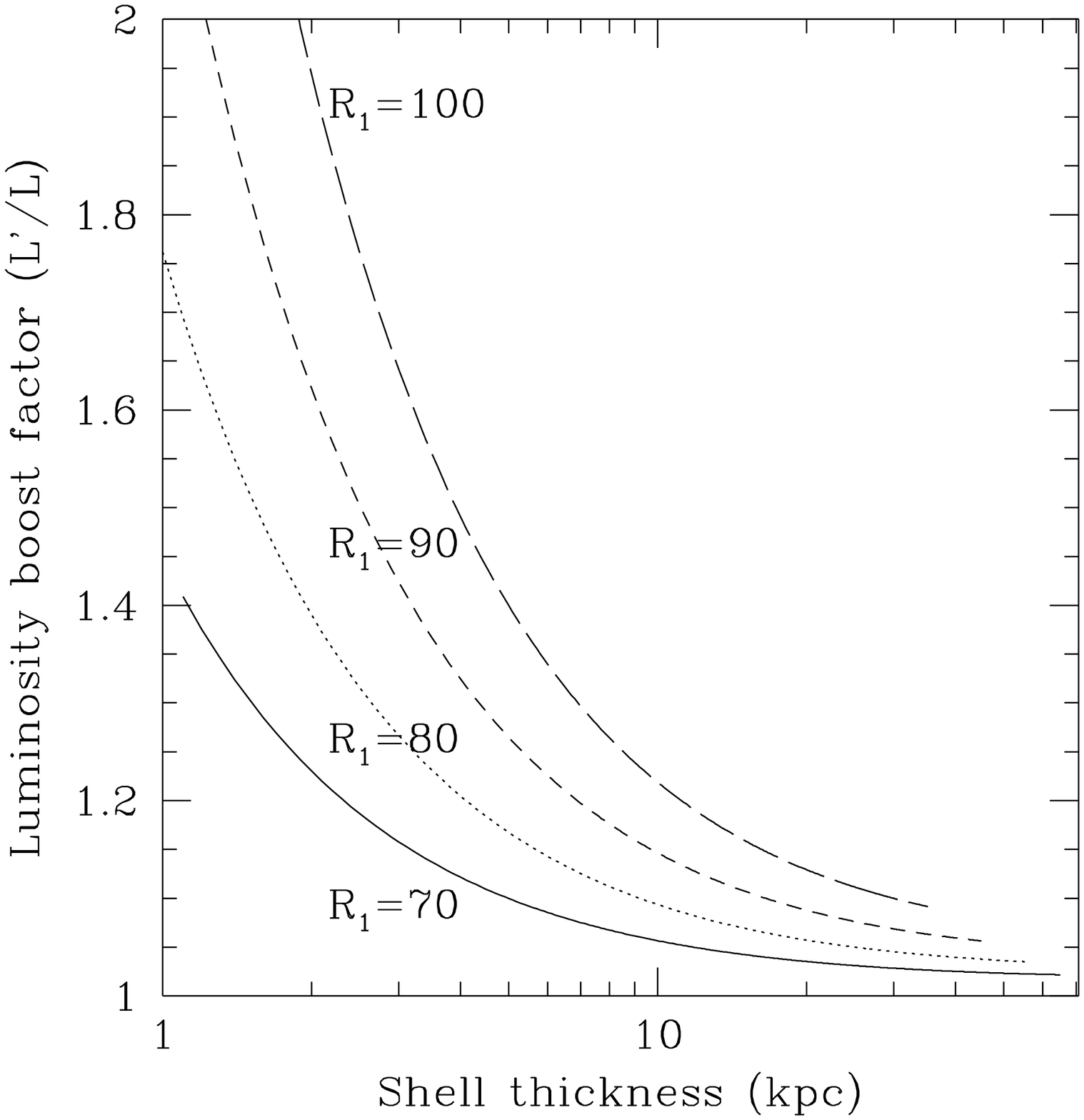}
\vspace{13cm}
\caption{
{\it (Top):} MOS1 image of MS0735 in the [0.3-10] keV energy band.
The image is corrected for vignetting and exposure.
The circular regions considered for the spectral analysis of the cavities
are indicated.
The N and S cavities have a radius of  $\sim 100$ kpc and
are located at a distance of $\sim 170$ and $\sim 180$
kpc from the cluster center, respectively.
{\it (Bottom, Left):} Geometry of the simplified phenomenological model 
considered in estimating the luminosity boost factor. 
During the cavity expansion, all the gas filling the cavities is assumed 
to be compressed into the bright shells. 
{\it (Bottom, Right):} 
Estimated luminosity boost factor due to cavity espansion and gas 
compression in the shells, as a function of the thickness of the shell
(see left panel for the geometry considered).
Different curves refer to different radius $R_1$ of the cavities 
(units in kpc). 
For an adiabatic index $\gamma = 5/3$, the maximum compression in a normal 
shock is a factor $f = 4$, so that the ratio of the shell thickness to the 
cavity radius is $R_2/R_1 - 1 > (1 + 1/f)^{1/3} - 1 \simeq 0.077$.
For the cavities observed in MS0735 ($R_1 \sim 100$ kpc), the shell thickness
is $\sim 8$ kpc, leading to a luminosity boost factor $\sim$ 25\%. 
}
\label{fig:1}       
\end{figure}
\begin{figure}[h]
\centering
\includegraphics[height=5.7cm]{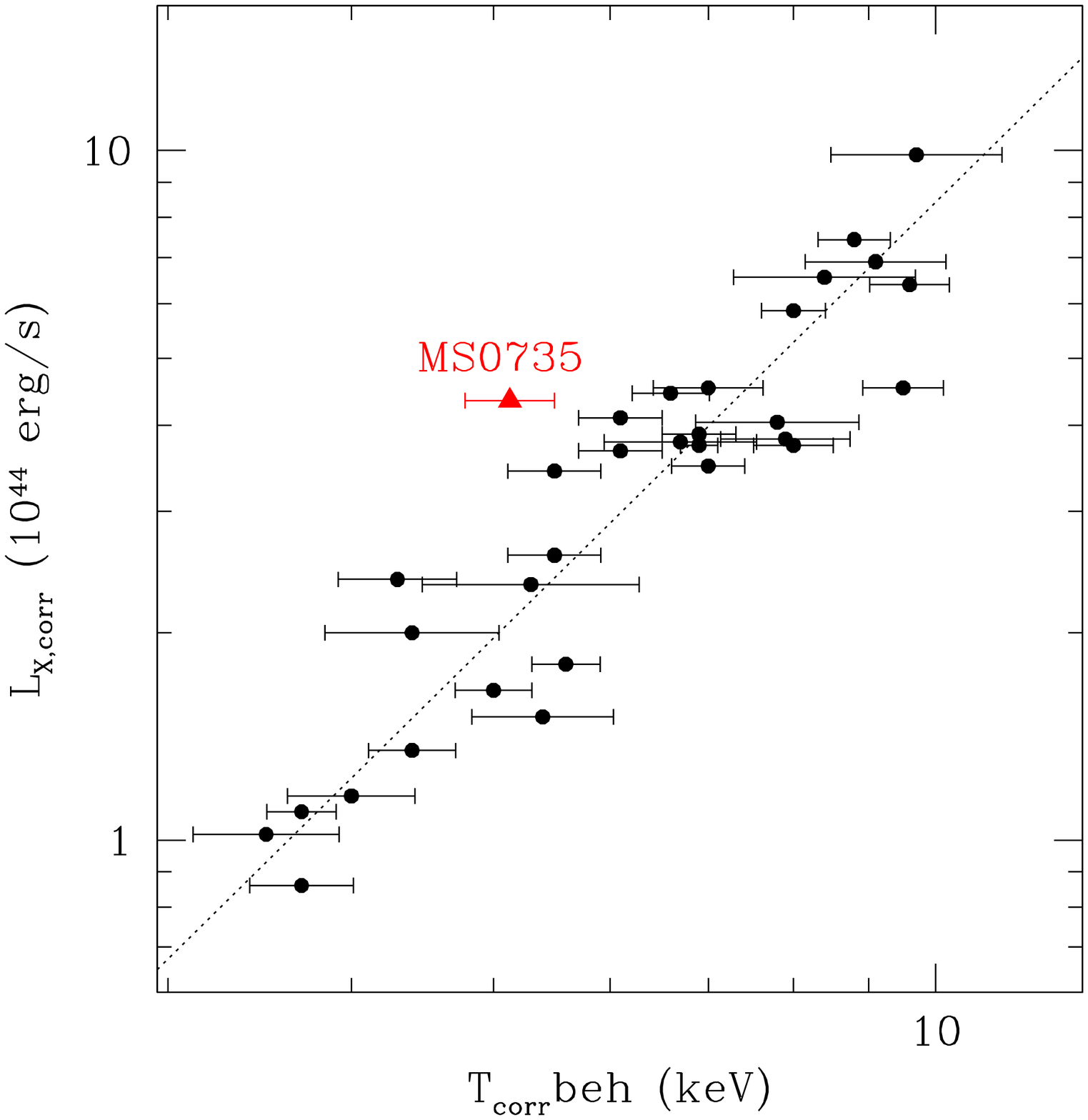}
\includegraphics[height=5.7cm]{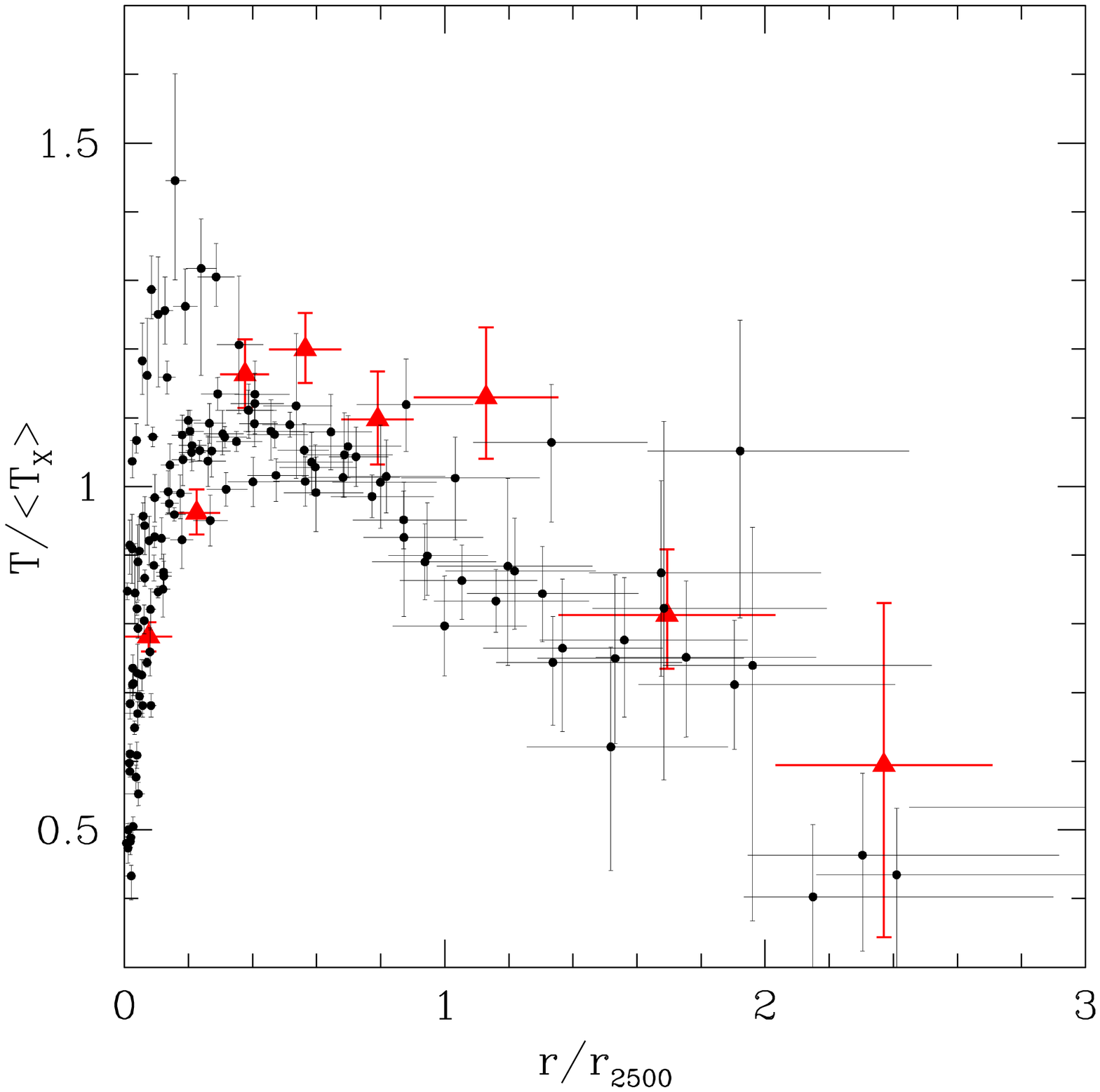}
\caption{
{\it (Left):} 
Observed X-ray luminosities (corrected for the effect of cooling flows
in the central 70 kpc)
vs. emission-weighted temperatures (derived excluding
cooling flow components) for a sample of nearby galaxy clusters 
\cite{markevitch98}.
The red triangle represents MS0735 data from the present observations.
Note that, since the cooling flow region in MS0735 is bigger than
the one adopted by \cite{markevitch98} for the cluster sample, 
the effect of cooling flow is not completely corrected. 
{\it (Right):}
Temperature profile measured for MS0735 (red triangles) overlaid onto the
temperature profile observed for a sample of 12 relaxed clusters 
\cite{vikhlinin05}.
The profiles for all clusters are projected and scaled in radial units of
$r_{2500}$. The temperatures are scaled to the cluster emission-weighted
temperature excluding the central 70 kpc regions.
}
\label{fig:2}       
\end{figure}

\item
The energetic outburst and the consequently rising cavities uplift the 
central cool, low-entropy gas up to large radii, and at the same time the 
compression in the shells increases the ICM density. 
This results in an increase of emissivity and thus luminosity.
An attempt to estimate the expected boost in luminosity is presented in 
Fig. \ref{fig:1} (bottom left), which shows a sketch of the 
simple phenomenological model of the structure and emissivity of the gas
considered to evaluate the effect of the cavity
expansion and compression of the ICM in the bright shells.
We estimate that the luminosity is boosted by a factor
which depends upon the cavity radius and shell thickness. 
For the particular configuration of the cavities observed in MS0735, we
expect an increase in luminosity by a factor of the order of about
25\% (see Fig. \ref{fig:1}, bottom right), consistent with our measurements.
The unabsorbed X-ray luminosity ([0.1-2.4] keV) estimated by the spectra 
extracted after excising the cooling flow region is
$\sim 3.8 \times 10^{44}$ erg/s.
To evaluate the effect of the cavities on the luminosity, this has to
be compared with the value estimated by the spectra extracted after 
excising the cavity regions (the cavities lie outside the cooling region),
which is $\sim 3.0 \times 10^{44}$ erg/s.
In both estimates the missing luminosity expected from a $\beta$-model 
profile inside the masked regions is added back in. 
The increased emissivity due to the cavities could also lead to an 
overestimate of the gas mass fraction, in qualitative agreement with the 
high value ($f_{\rm gas, 2500}= 0.165 \pm 0.040$) that we measure for MS0735. 
\\

\item
MS0735 is a factor $\sim$2 more luminous than expected from its average 
temperature on the basis of the observed $L$-$T$ relation for galaxy
clusters (\cite{markevitch98}, see Fig. \ref{fig:2}, left).
This effect may be partially explained by the boost in luminosity
due to the cavities.
Besides this, no obvious immediate impact on properties such as the 
scaled temperature profile (see Fig. \ref{fig:2}, right)
and scaled metallicity profile show up from our analysis. 
Also, the quantities we measure for MS0735 are consistent with the 
$M$-$T$ relation predicted by the cluster scaling laws.  
We conclude that violent outbursts such as the one in MS0735 do not cause gross 
instantaneous departures from cluster scaling relations 
(other than the $L$-$T$ relation).  
However, if they are relatively common they may play a role in shaping these
relations.

\end{itemize}

%%%%%%%%%%%%%%%%%%%%%%%%%%%%%%%%%%%%%%%%%%%%%%%%%%%%%%%%%%%%%%%%%%%%%%%%%%%%%

\subsubsection*{Acknowledgments}

We thank A. Vikhlinin for sending the data
used to make the plots in Fig. \ref{fig:2}, right.
This research is supported by NASA grant NNG05GK876 and by NASA Long Term
Space Astrophysics Grant NA64-11025.

%%%%%%%%%%%%%%%%%%%%%%%%%%%%%%%%%%%%%%%%%%%%%%%%%%%%%%%%%%%%%%%%%%%%%%%%%%%%%

%
%
% BibTeX users please use
% \bibliographystyle{}
% \bibliography{}
%
% Non-BibTeX users please follow the syntax
% the syntax of "referenc.tex" for your own citations
%%%%%%%%%%%%%%%%%%%%%%%% referenc.tex %%%%%%%%%%%%%%%%%%%%%%%%%%%%%%
% sample references
% "physics"
%
% Use this file as a template for your own input.
%
%%%%%%%%%%%%%%%%%%%%%%%% Springer-Verlag %%%%%%%%%%%%%%%%%%%%%%%%%%

%
% BibTeX users please use
% \bibliographystyle{}
% \bibliography{}

%%%%%%%%%%%%%%%%%%%%%%%%%%%%%%%%%%%%%%%%%%%%%%%%%%%%%%%%%%%%%%%%%%%%%%  }

%%%%%%%%%%%%%%%%%%%%%%%%%%%%%%%%%%%%%%%%%%%%%%%%%%%%%%%%%%%%%%%%%%%%%%

\printindex
\end{document}